\documentclass[fleqn,usenatbib]{mnras}

\usepackage{newtxtext,newtxmath}

\usepackage[T1]{fontenc}

\DeclareRobustCommand{\VAN}[3]{#2}
\let\VANthebibliography\thebibliography
\def\thebibliography{\DeclareRobustCommand{\VAN}[3]{##3}\VANthebibliography}

\usepackage{graphicx}	
\usepackage{amsmath}	
\usepackage{wasysym}

\title[GPI HCI for HD 29992 and HD 196385]{High-contrast imaging of HD 29992 and HD 196385 with GPI}

\author[L. H. Garc\'ia et al.]{Luciano H. Garc\'ia,$^{1}$\thanks{E-mail: luciano.garcia.030@unc.edu.ar}
R. Petrucci,$^{1,2}$
E. Jofr\'e,$^{1,2,3}$
and M. G\'omez$^{1,2}$
\\
$^{1}$Universidad Nacional de C\'ordoba, Observatorio Astron\'omico de C\'ordoba, Laprida 854, X5000BGR C\'ordoba, Argentina\\
$^{2}$Consejo Nacional de Investigaciones Cient\'ificas y T\'ecnicas (CONICET), Godoy Cruz 2290, CABA, CPC 1425FQB, Argentina\\
$^{3}$Instituto de Astronom\'ia, Universidad Nacional Aut\'onoma de M\'exico, Circuito Exterior, C.U., A. Postal 70-264, 04510 Ciudad de M\'exico, M\'exico\\
}

\date{Accepted XXX. Received YYY; in original form ZZZ}

\pubyear{2022}

\begin{document}
\label{firstpage}
\pagerange{\pageref{firstpage}--\pageref{lastpage}}
\maketitle

\begin{abstract}
Based on high contrast images obtained with the Gemini Planet Imager (GPI), we report the discovery of two point-like sources at angular separations of $\rho\sim0.18\arcsec$~and $\rho\sim0.80\arcsec$~from the stars HD 29992 and HD 196385. A combined analysis of the new GPI observations and images from the literature indicates that the source close to HD 29992 could be a companion to the star. Concerning HD 196385, the small number of contaminants ($\sim0.5$) suggests that the detected source may be gravitationally bound to the star. For both systems, we discarded the presence of other potential companions with $m>75$ M$_{\rm Jup}$ at $\rho\sim0.3 - 1.3\arcsec$.~From stellar model atmospheres and low-resolution GPI spectra, we derive masses of $\sim0.2$ - $0.3$ M$_{\odot}$ for these sources. Using a Markov-chain Monte Carlo approach, we performed a joint fit of the new astrometry measurements and published radial velocity data to characterize the possible orbits. For HD 196385B, the median dynamic mass is in agreement with that derived from model atmospheres, whilst for HD 29992B, the orbital fit favors masses close to the brown dwarf regime($\sim0.08$ M$_{\odot}$). HD 29992 and HD 196385 might be two new binary systems with M-type stellar companions. However, new high angular resolution images would help to definitively confirm whether the detected sources are gravitationally bound to their respective stars, and permit tighter constraints on the orbital parameters of both systems.
\end{abstract}




\section{Introduction}
\label{sec:intro}

The direct imaging technique has made possible the discovery of tens of wide ($a\geq10$ au), low-mass companions ($m<75$ M$_{\rm Jup}$) that populate a little explored region of the mass vs. orbital radius diagram \citep[see][for a review]{bowler2016}. Nevertheless, despite these findings, their occurrence rate is still uncertain \citep{launhardt2020}. In addition, several imaged companions have been detected inside the gap of debris disks, such as HR 2562B \citep{konopacky2016} and HD 193571B \citep{barcucci2019}, with these discoveries having challenged our current understanding about the formation and later evolution of these companions. Monitoring direct imaged companions allows their dynamic mass to be estimated (which is independent of the assumptions of evolutionary models), so they can be used as empirical benchmarks for evolutionary models, especially of low-mass objects \citep{franson2022}. Hence, there is a need for increasing the number of known systems with wide, low-mass companions, in order to provide a better estimate of the rate of occurrence and an improved understanding of the predominant mechanisms that shape such systems.\\

Direct imaging combined with coronagraphy is currently one of the most efficient techniques for searching for companions in wide orbits. Here, we used the Gemini Planet Imager at Gemini South \citep[GPI,][]{macintosh2014} to obtain high contrast, and high angular resolution coronagraphic images of two stars with evidence of wide, low-mass companions through radial velocity monitoring: HD 29992, and HD 196385. In this work, we report the discovery of two point-like sources in the high contrast images of these stars obtained with GPI. In Section~\ref{sec:tar}, we present the targets and briefly review previous studies that have searched for companions around both stars. The observations and the data reduction are described in Section~\ref{sec:obs}. In Section~\ref{sec:resul} we investigate the companionship and characteristics of the detected sources. To put some initial constraints on the possible orbits, we used the package \texttt{orbitize!} to perform a joint fit of the new astrometry data from GPI and existing radial velocity measurements from the literature. Finally, in Section \ref{sec:conc}, we summarize the results and present our conclusions.\\

\section{Properties of the targets and previous surveys for companions}
\label{sec:tar}

\begin{table*}
\centering
  \caption{Relevant parameters for the observed targets.}
  \label{t:1a}
 \begin{tabular}{lcccccccc}
    \hline
    Target &\multicolumn{1}{c}{$V^{\star}$} & \multicolumn{1}{c}{$H^{\spadesuit}$} & \multicolumn{1}{c}{$d^{\dag}$} & \multicolumn{1}{c}{$\mu_{\alpha}\times cos(\delta)^{\dag}$} & \multicolumn{1}{c}{$\mu_{\delta}$$^{\dag}$} & \multicolumn{1}{c}{SpT$^{\star}$} & \multicolumn{1}{c}{mass$^{\ddagger}$} &\multicolumn{1}{c}{Age$^{\blacklozenge}$} \\
           & (mag)                 & (mag)                  & (pc)                      & (mas)                                                  & (mas)                                &                  &	(M$_{\odot}$) &(Gyr)            \\                
	\hline
    HD 29992  &  5.1 & 4.3 & $28.8\pm 0.1$ & $42.56\pm0.06$ & $212.70\pm0.07$ & F3IV & $1.47\pm 0.24$ &$1.7^{+0.2}_{-0.2}$ \\
	HD 196385 &  6.4 & 5.6 & $47.7\pm0.1$ & $73.04\pm0.02$ & $-16.41\pm0.02$ & A9V & $1.53\pm 0.27$ &$1.2^{+0.3}_{-0.7}$ \\
    \hline
	\multicolumn{8}{l}{$^{\star}$\textit{SIMBAD} database}\\
	\multicolumn{8}{l}{$^{\spadesuit}$2MASS All Sky Catalog of point sources \citep{cutri2003}}\\
	\multicolumn{8}{l}{$^{\dag}$\textit{GAIA} DR2 catalog \citep{gaia2018}}\\
	\multicolumn{8}{l}{$^{\ddagger}$\citet{stassun2019}}\\
	\multicolumn{8}{l}{$^{\blacklozenge}$\citet{casagrande2011}}\\
  \end{tabular}
\end{table*}

We selected the stars HD 29992 and HD 196385, because both have shown evidence of low-mass companions in wide orbits. \citet{borgniet2016} studied the long term variations of the radial velocity (RV) of these stars, and for HD 29992 they found a quadratic RV trend over the $\sim4.8$ year time baseline of their compiled data. Taking into account this time baseline and also the amplitude of the variations, the authors inferred the presence of a companion, most likely a low-mass star (despite them estimating a minimum mass of $\sim40$ M$_{\rm Jup}$) at a separation larger than 3 au. For HD 196385, \citet{borgniet2016} reported a linear trend in the RV observations, but were not able to provide a limit on the separation or on the mass of a potential companion. Basic parameters of the stars are listed in Table \ref{t:1a}.\\

Previous attempts to detect sub-stellar mass companions around HD 29992 through high contrast, high angular resolution images were carried out by \citet{ehrenreich2010} using NaCo/VLT. These authors obtained direct and coronagraphic (with a Focal Plane Mask of diameter $\invdiameter=0.7\arcsec$) images, but did not report any sources in the images and discarded the presence of companions with masses $m>75$ M$_{\rm Jup}$ beyond $\sim1.9\arcsec$ ($\sim55$ au) from the central star. For HD 196385, we are not aware of any previous attempts to detect companions through high angular resolution images.\\

\section{Observations and data reduction}
\label{sec:obs}
High contrast images and low-resolution spectra ($R\sim45$) in the H-band for any source in the FOV ($\sim2.8\times2.8\arcsec$) around the selected stars HD 29992 and HD 196385 were obtained using the Integral Field Spectroscopy mode (IFS; spatial scale 14.13 mas/lenslet) of GPI in combination with a Focal Plane Mask (FPM; diameter $\invdiameter=246$ mas). The observations for HD 29992 were executed in November 2019, and 24 images were obtained for this target. HD 196385 was observed in September 2019, but only 4 images were obtained (program GS-2019B-Q-107; PI: Luciano H\'ector Garc\'ia). All the observations were carried out in the pupil tracking mode\footnote{In this mode, the FOV is allowed to rotate around the central star.} in order to apply the Angular Differential Imaging (ADI) technique \citep{marois2006}.\\

For processing the raw data and constructing the 3D wavelength calibrated datacubes ($x,y,\lambda$), we used the GPI Data Reduction Pipeline\footnote{Available at: http://docs.planetimager.org/pipeline/} version 1.3.0 \citep[DRP;][]{perrin2016}. The usual primitives\footnote{A primitive is the name that an elementary algorithm (to be applied on the files for example, for removing the dark background, or building the 3D datacubes) receives in the GPI-DRP.} were applied for subtracting the dark background, correcting for variations between the CCD readout channels and microphonics, interpolating over bad pixels, compensating for flexure, and converting from raw 2D IFS frames to 3D datacubes. Argon arc lamp exposures taken prior to the science exposure sequences were used for wavelength calibration, and to determine the locations and fluxes of the four satellite spots created by the apodizer. The satellite spots were used for image registration. The limited number of images (and sky rotation) for HD 196385 did not allow us to further process the images to remove the PSF of the star and instrumental scattered light. However, in the case of HD 29992, it was possible to apply the KLIP algorithm (Karhunen-Lo\`eve Image Projection) provided by the Python library pyKLIP\footnote{Available at: https://pyklip.readthedocs.io/en/latest/index.html} \citep{wang2015} to correct the final image for these effects. For the astrometric calibrations, we adopted a platescale of $14.166\pm0.007$ mas px$^{-1}$ \citep{konopacky2014,derosa2015}. This value is consistent with a more recent measurement of the pixel scale ($14.161\pm0.021$ mas px$^{-1}$) obtained by \citet{derosa2020}. The measured Position Angles (PAs) were adjusted by $0.45\pm0.11$ deg according to the updated north offset angle given by \citet{derosa2020}, and we used the offset closest to the date of the observations from this study.\\

\section{Analysis and results}
\label{sec:resul}

\begin{figure}
  \includegraphics[width=0.49\linewidth,trim={3cm 0cm 3cm 0cm},clip]{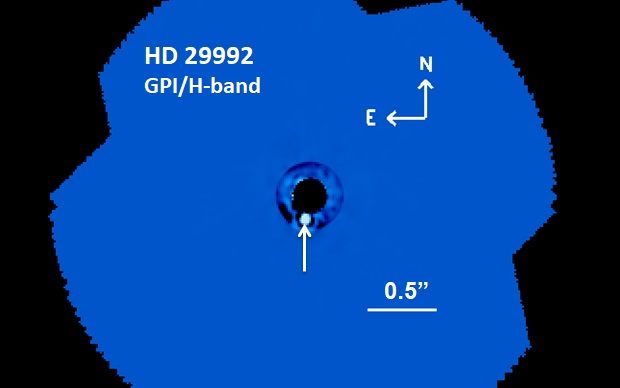}
  \includegraphics[width=0.49\linewidth,trim={3cm 0cm 3cm 0cm},clip]{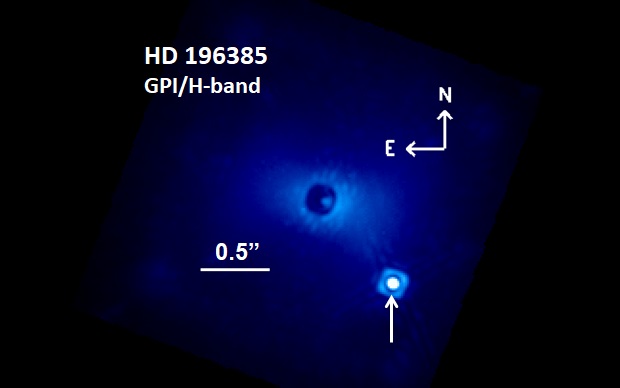}
  \caption{\emph{Left:} ADI combination of the 24 images for HD 29992 using the KLIP algorithm. The wavelengths of the datacube are median-combined. A point-like source is clearly visible towards the south-east of the star. \emph{Right:} Median combined image for one of the datacubes of HD 196385. The source at $\rho\sim0.8\arcsec$ from the star is distinctly present.}
  \label{fig:1}
\end{figure}

\begin{table}
	\centering
	\caption{Measured parameters for the newly detected sources.$^{*}$}
	\label{t:1}
	\begin{tabular}{lcccc} 
		\hline
		    Target &\multicolumn{1}{c}{$\rho$} & \multicolumn{1}{c}{Separation} & \multicolumn{1}{c}{PA} & \multicolumn{1}{c}{$\Delta H$} \\
		    &\multicolumn{1}{c}{($\times 10^{-3}$\arcsec)} & \multicolumn{1}{c}{(au)} & \multicolumn{1}{c}{($\degr$)} & \multicolumn{1}{c}{(mag)} \\
		\hline
    HD 29992B  & $175\pm 1$ & $5.04\pm0.04$ & $168.1\pm0.3$ & $5.9\pm0.5$ \\
	HD 196385B & $802\pm 1$ & $38.3\pm0.1$ & $219.1\pm0.2$ & $4.9\pm0.2$ \\
		\hline
    \multicolumn{5}{l}{$^{*}$We are using the designation HD 29992B and HD 196385B}\\
    \multicolumn{5}{l}{for simplicity here.}\\
	\end{tabular}
\end{table}

\begin{figure}
  \includegraphics[width=0.49\linewidth,trim={0.5cm 0.5cm 0.5cm 0.5cm},clip]{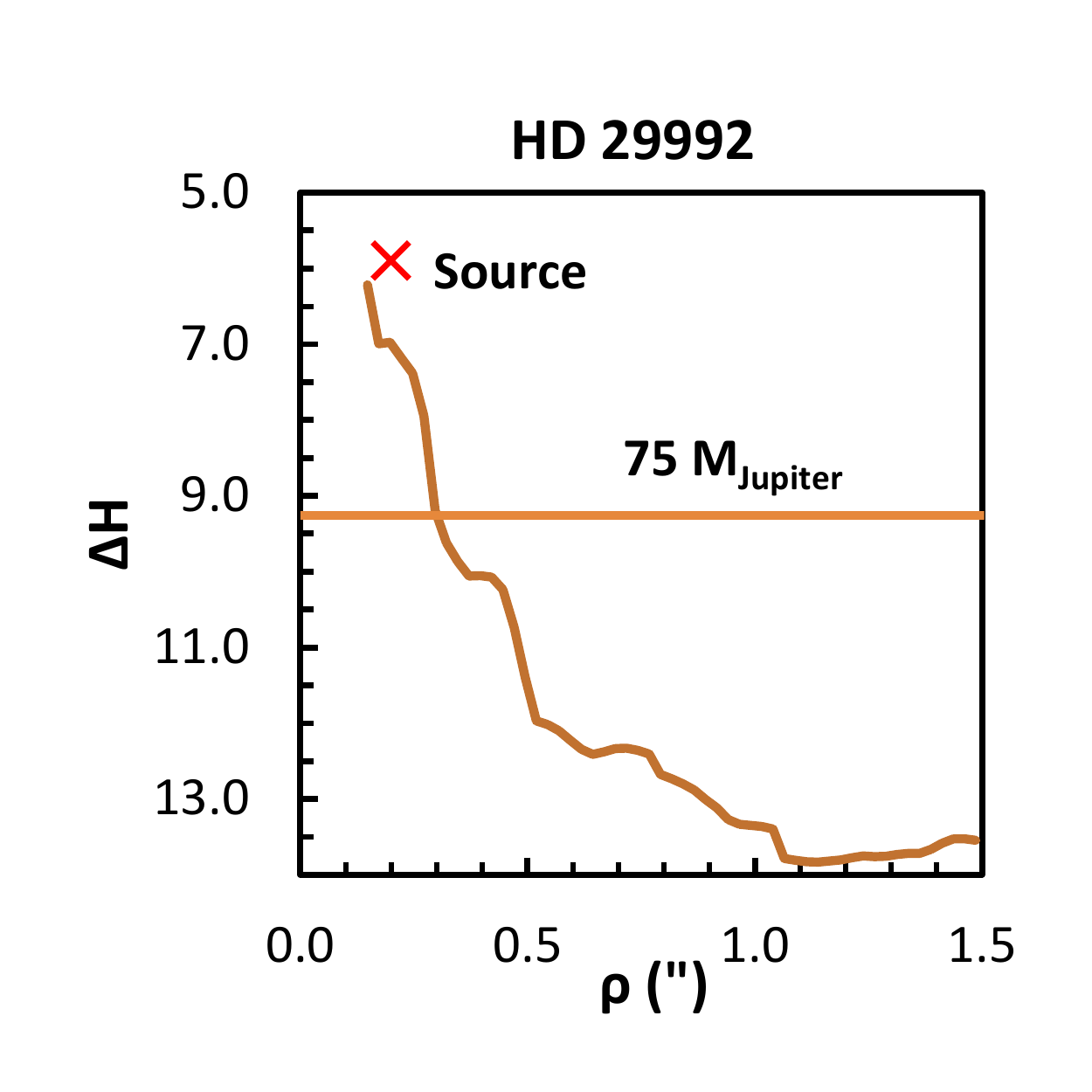}%
  \hfill
  \includegraphics[width=0.49\linewidth,trim={0.5cm 0.5cm 0.5cm 0.5cm},clip]{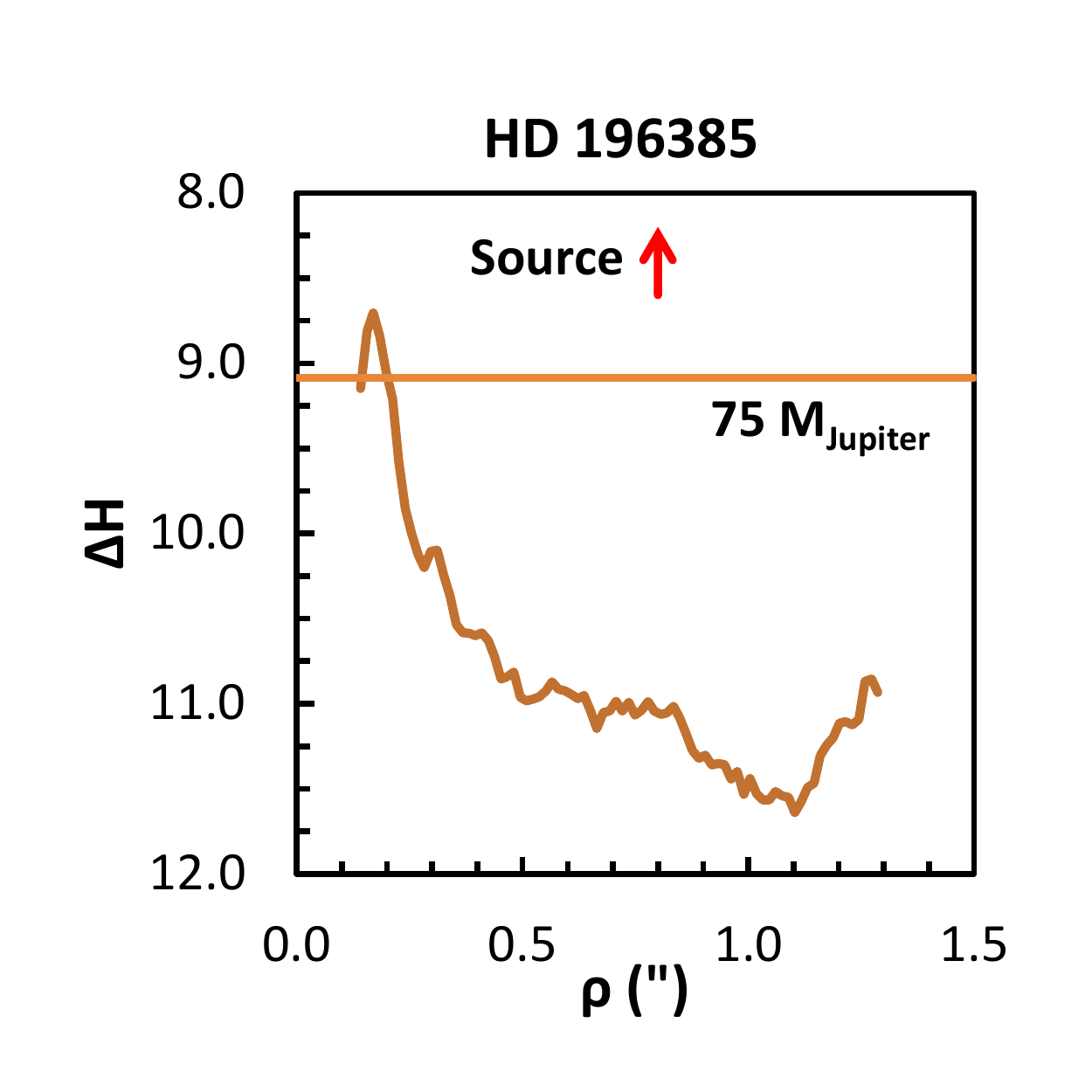}
  \caption{\emph{Left:} Measured contrast curve (algorithm throughput corrected) for the final image of HD 29992. The horizontal line represents the contrast of a companion with a mass $m=75$ M$_{\rm Jup}$ according to the models of \citet{baraffe2003}. The cross shows the location of the detected source. \emph{Right:} Measured contrast for HD 196385. The arrow indicates that the contrast of the detected source falls outside the range of the plot. The horizontal line represents the contrast of a 75 M$_{\rm Jup}$ companion.}
  \label{fig:2}
\end{figure}

\begin{figure}
  \includegraphics[width=0.49\linewidth,trim={0.5cm 0.5cm 0.5cm 0.5cm},clip]{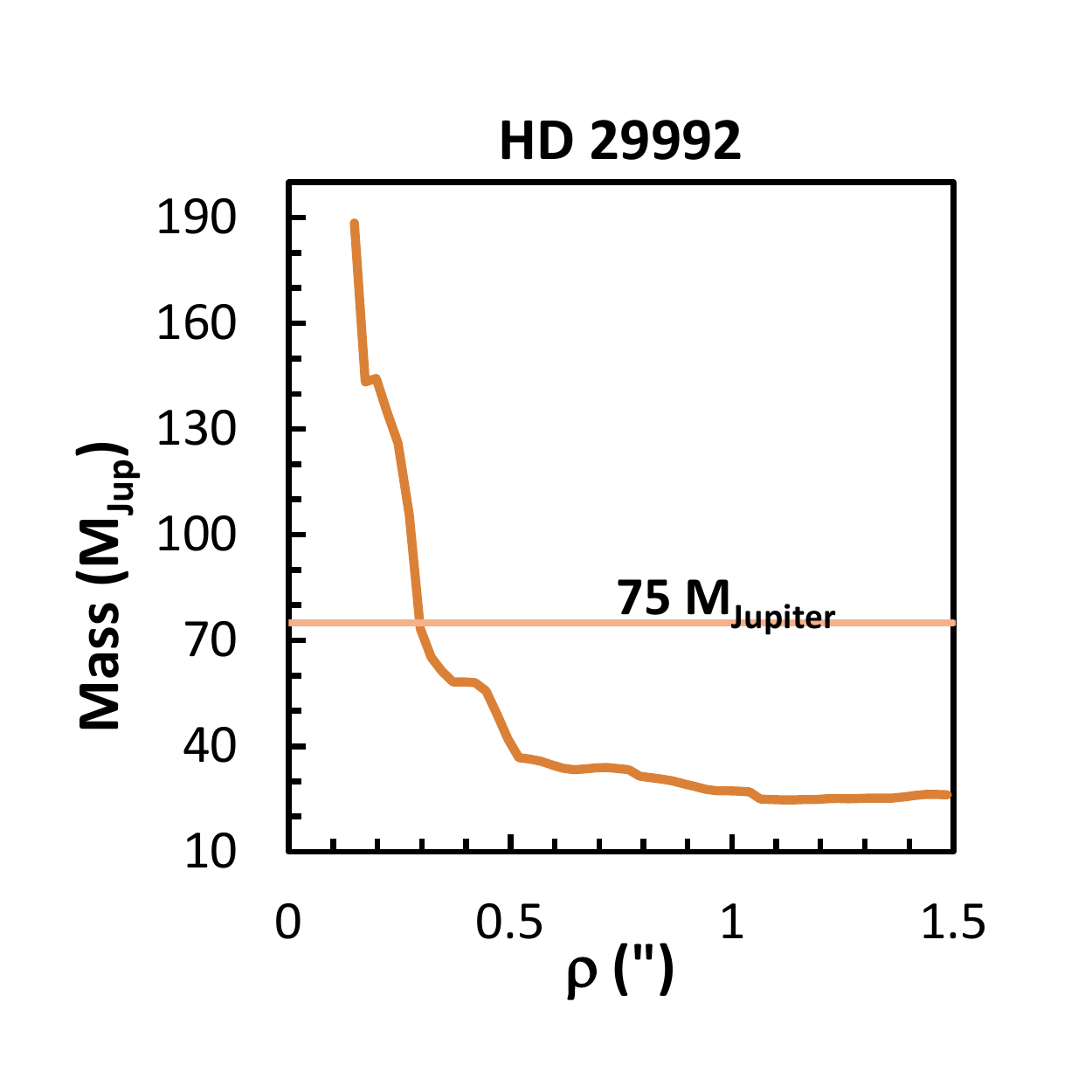}
  \includegraphics[width=0.49\linewidth,trim={0.5cm 0.5cm 0.5cm 0.5cm},clip]{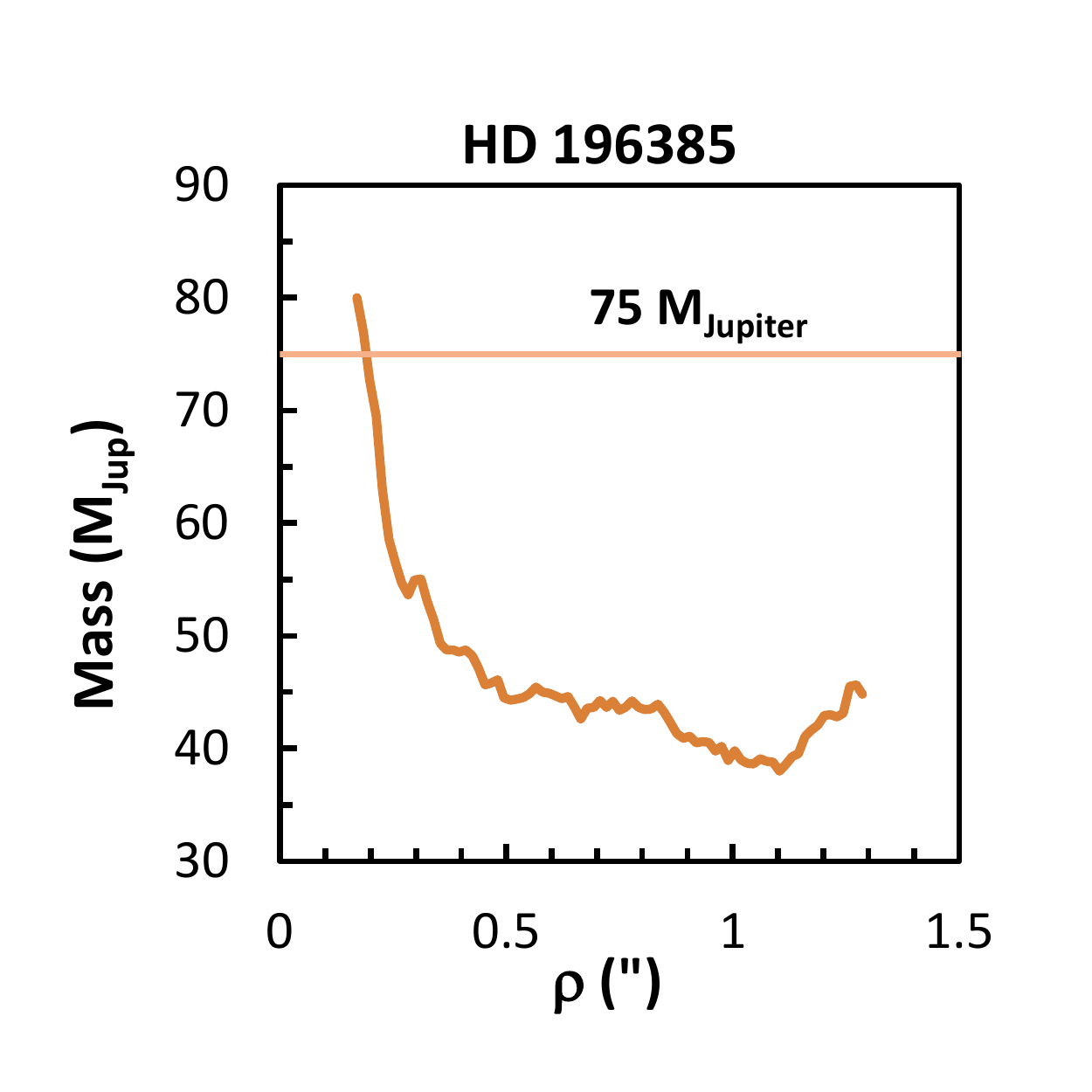}
  \caption{Mass limits calculated from combining the measured contrasts and the models of \citet{baraffe2003}.~Any companion in the FOV of GPI with a mass above the curves would have been detected in the images. Horizontal lines indicate the location of a companion with a mass $m\sim75$ M$_{\rm Jup}$.}
  \label{fig:3}
\end{figure}

Figure~\ref{fig:1} shows the GPI images for HD 29992 and HD 196385 after applying the reduction steps listed in section~\ref{sec:obs}. A point-like source is visible in the images of each star, with both sources being bright enough to be identifiable in the individual datacubes. The angular ($\rho$) and projected separations, plus position angles (PA) and measured contrasts ($\Delta H$), are listed in Table~\ref{t:1}. When processing the raw data, we also extracted the achieved contrast vs. angular separation to the central star. In the case of HD 29992, the contrast was corrected from the throughput of the ADI algorithm, with the resulting curves shown in Figure~\ref{fig:2}. We combined the contrast curves with models of atmospheres for sub-stellar mass objects to obtain the approximate mass limits to the companions that could be detected in the images. Figure~\ref{fig:3} displays these resulting limits using the models from \citet{baraffe2003}. In addition to the mentioned sources, we can discard the presence of companions with mass $m>75$ M$_{\rm Jup}$ beyond $\rho\sim0.3\arcsec$ ($\sim8$ au) and $\rho\sim0.2\arcsec$ ($\sim9$ au) orbiting around HD 29992 and HD 196385, respectively. Also, by taking into account the results from \citet{ehrenreich2010}, it is possible to discard the presence of companions with $m>75$ M$_{\rm Jup}$ between $\rho\sim0.3\arcsec$ and $\rho\sim10\arcsec$ for HD 29992.\\

\subsection{Association with the central star}

The search for companions through images must address the possibility of a fortuitous alignment with stars projected in the FOV, especially for those at wide separations. A common practice is to use multi-epoch images for comparing the proper motion of a potential companion with that of the central star, in order to infer whether both are co-moving in the sky. However, GPI was unfortunately taken out of service during the first half of 2020, before we were able to obtain 2nd-epoch images to measure the proper motions of the detected sources. Therefore, in this section, we discuss other ways of examining whether the detected sources may be gravitationally bound to their respective central stars.\\

\subsubsection{HD 29992}
\label{sec:comphd29992}

\begin{figure}
  \includegraphics[width=0.9\columnwidth,trim={1cm 1cm 1cm 1cm},clip]{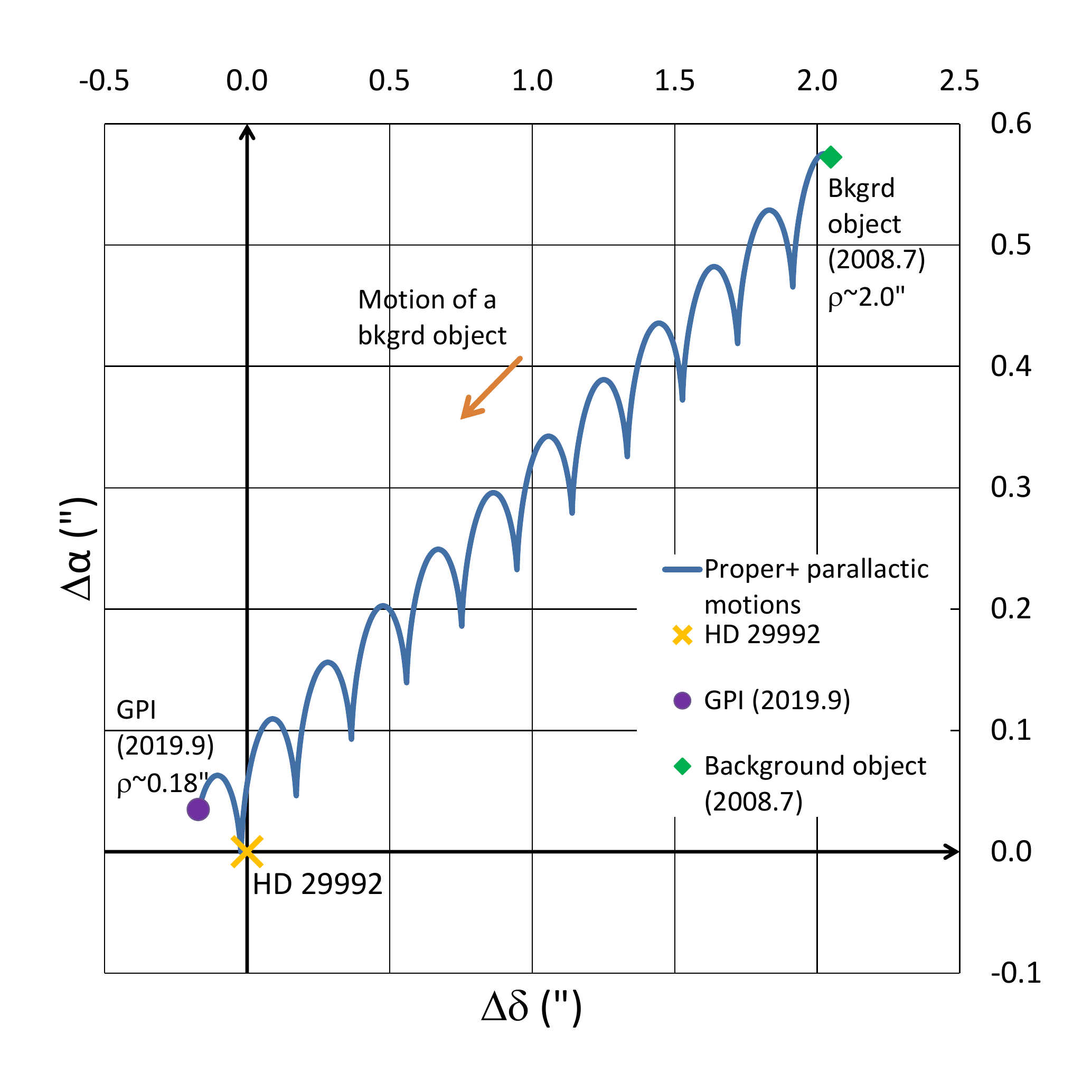}
  \caption{Relative position of the detected source with respect to HD 29992 (orange cross). The purple circle indicates its location in the GPI images (date $=2019.9$). The green square is the calculated position at the date of the~\citet{ehrenreich2010} images (2008.7), assuming that the source is an infinitely distant object.}
  \label{fig:4}
\end{figure}

For HD 29992, it was possible use the images from \citet{ehrenreich2010} to evaluate the companionship of the source detected in the GPI images. By assuming that this is an infinitely distant object, we used the proper motion of the star and the position in the GPI images to calculate its approximate angular separation and position angle on the date (2008.7) of the images from \citet{ehrenreich2010}. We named this date Epoch 1. A background object will show an opposite drift in comparison to the proper motion of the star. The GPI images (Epoch 2 - 2019.9) are separated by just over 11 years from Epoch 1. This relatively large time baseline and the high proper motion of the star allowed us to ignore sources of error in the proper motion of the star and the measured position of the potential companion. At Epoch 1, we calculated that the approximate angular separation of the source would be $\rho\sim2\arcsec$ (PA~$\sim13.1\degr$), assuming that this is an infinitely distant object. This is illustrated in Figure~\ref{fig:4}. An object with the contrast observed in the GPI images, located at $\rho\sim2\arcsec$ from HD 29992 in Epoch 1, should have been detected in the \citet{ehrenreich2010} images. This lack of detection suggests that the source is co-moving with the star and that it was hidden by the FPM, or that it fell in the glare of the central star in the \citet{ehrenreich2010} images.\\

As mentioned in section \ref{sec:tar}, \citet{borgniet2016} estimated that a companion around HD 29992 must be located at a separation larger than 3 au, and is most likely a low-mass star. This is in good agreement with the separation measured from the GPI images ($\sim5$ au) and the estimated mass of $m\sim0.2$ M$_{\odot}$ (see sections \ref{characterization} and \ref{orbits}). To address the question of whether the \citet{borgniet2016} observations could have detected the source seen in the GPI images, we calculated the inclination $i_{\star}$ of the system. To carry this out, it was necessary to determine the rotation period of the star for the first time. We used the data from the Transiting Exoplanet Survey Satellite \citep[TESS;][]{Ricker2015}. The details of this calculation are presented in Appendix \ref{sec:ap-A}. The value obtained for $i_{\star}$ was $\sim65.3\degr$. According to the sensitivity limits from \citet[][see Figure 9 from their work]{borgniet2016} and the resulting $i_{\star}$, a companion with a mass of $m\sim0.2$ M$_{\odot}$ located at $\sim5$ au from the star should have been detected in their observations. In addition, there is no evidence for other companions, pointing to the source detected in the images with GPI as being the possible origin for the trend observed in the RV of the star. Therefore, the trend found by \citet{borgniet2016}, plus the 1st-epoch observations from \citet{ehrenreich2010}, lead us to propose that the source in the images of HD 29992 is gravitationally bound to the star. \\

\subsubsection{HD 196385}
\label{sec:comphd196385}

\begin{table}
\centering
  \caption{Adopted luminosity function.}
  \label{t:2}
  \setlength{\tabcolsep}{3pt}
 \begin{tabular}{lcc|lcc}
    \hline
    \multicolumn{1}{c}{SpT} & \multicolumn{1}{c}{$M_{H}$} & \multicolumn{1}{c|}{Number} & \multicolumn{1}{c}{SpT} & \multicolumn{1}{c}{$M_{H}$} & \multicolumn{1}{c}{Number}\\
	                        & (mag)                          & (pc$^3$/SpT/$1\times10^{-3}$) &                       &(mag)                           & (pc$^3$/SpT/$1\times10^{-3}$) \\ 
    \hline
B8	&	0.1	&	0.04	&	K2	&	4.4	&	3.10	\\
A0	&	0.6	&	0.30	&	K4	&	4.6	&	2.50	\\
A2	&	1.2	&	0.40	&	K5	&	4.6	&	2.50	\\
A5	&	1.5	&	0.50	&	K7	&	4.9	&	4.50	\\
A7	&	1.7	&	0.50	&	M0	&	5.4	&	5.00	\\
F0	&	2.0	&	0.70	&	M1	&	5.7	&	5.00	\\
F2	&	2.2	&	0.70	&	M2	&	6.1	&	5.20	\\
F5	&	2.7	&	0.80	&	M3	&	6.7	&	11.00	\\
F8	&	3.1	&	1.50	&	M4	&	7.5	&	9.50	\\
G0	&	3.3	&	1.90	&	M5	&	8.5	&	7.00	\\
G2	&	3.4	&	1.90	&	M6	&	9.5	&	5.00	\\
G5	&	3.6	&	1.90	&	M7	&	10.1	&	3.00	\\
G8	&	3.9	&	2.60	&	M8	&	10.5	&	2.00	\\
K0	&	4.1	&	2.60	&	M9	&	10.7	&	2.00	\\
    \hline
  \end{tabular}
\end{table}

For HD 196385, we used statistical arguments to evaluate the chances of finding a field star in the images with GPI by considering how many B8-M9 type stars would be expected in the GPI FOV. The spectral types taken into account were limited to those that contributed the most to the number of potential contaminants due to their spatial density and detection distances. The calculation requires knowledge of the volume density and luminosity of the field stars. We assumed an isotropic distribution, and neglected the decrease of volume density with increasing distance from the Galactic plane, although this approximation will overestimate the number of potential contaminants. For the luminosity function, we combined the results in the J-band from \citet{cruz2003} and \citet{reid2004}. Also, the magnitude-SpT relation from \citet{kraus2007} and the absolute magnitudes from \citet{kirkpatrick2012} were used to calculate the corresponding H-band luminosity function. \\ 

Next, we calculated the maximum distances at which our images may detect each spectral type, assuming a uniform sensitivity between $\rho\sim0.1\arcsec$~(at the edge of the FPM) and the angular separation of the detected source, and using the highest sensitivity achieved in the images of each target. By making this approximation, we are overestimating the capability of detecting potential contaminants. Then, the corresponding volume was calculated and multiplied by the density listed in Table \ref{t:2}. By adding the results from all the spectral types in Table \ref{t:2}, we found a total of $\sim0.5$ objects per GPI field. It is worth mentioning that the actual number of contaminants is likely to be smaller than the one calculated here, as we have not only ignored the decrease in density when moving away from the Galactic plane, but we have also optimized the ability to detect contaminants by using the maximum sensitivity for all angular distances.\\

To check the result obtained by this simple method, we simulated a field using the area covered by GPI ($\sim7\times10^{-7}$ deg$^{2}$) centered on the galactic coordinates of HD 196385 ($l=293.1\degr$, $b=-33.7\degr$) using the Trilegal galactic model online tool\footnote{Available at: http://stev.oapd.inaf.it/cgi-bin/trilegal} \citep{girardi2012}. The default parameters were used for the bulge, halo, thin/thick disks, and the lognormal initial mass function from \citet{chabrier2001}. The limiting magnitude in the H-band was set to $\sim17$, according to the achieved contrast. Then, the number of objects calculated by the tool was zero, most likely due to the extremely small area covered by GPI. Thus, the low number of expected contaminants per image suggests that the chance of a fortuitous alignment is rather low, and that the detected source may be an actual companion to HD 196385.\\

\citet{borgniet2016} also studied the long-term RV variations of HD 196385. Following the same procedure as with HD 29992, we calculated for the first time the rotation period of HD 196385 using TESS photometry, obtaining $i_{\star}\sim16.9\degr$. A companion with $m\sim0.3$ M$_{\odot}$ (see section \ref{characterization}), at $\sim38$ au from the star could not have been detected in the observations of \citet[][see Figure 9 from their work]{borgniet2016}. However, the authors reported a linear trend in the radial velocity that could have been due to a companion to HD 196385, but they were not able to estimate a limit for its mass or separation. Therefore, it is difficult to validate whether this companion corresponded to the source identified in the GPI images. \\

\subsection{Determination of spectral types for the companion stars}
\label{characterization}

\begin{table}
\centering
  \caption{Absolute magnitude, mass, and effective temperature of the detected sources.}
  \label{t:3}
 \begin{tabular}{lccc}
    \hline
    Target &\multicolumn{1}{c}{$M_{H}$ (mag)} & \multicolumn{1}{c}{mass (M$_{\odot}$)} & \multicolumn{1}{c}{Teff (K)}  \\
    \hline
    HD 29992B  & $7.9\pm0.6$ & $0.19\pm0.05$ & $3593\pm160$ \\
	HD 196385B & $7.1\pm0.2$ & $0.28\pm0.03$ & $3790\pm24$ \\
    \hline
  \end{tabular}
\end{table}

\begin{figure}
  \includegraphics[width=0.49\linewidth,trim={0.5cm 0.8cm 0.8cm 1cm},clip]{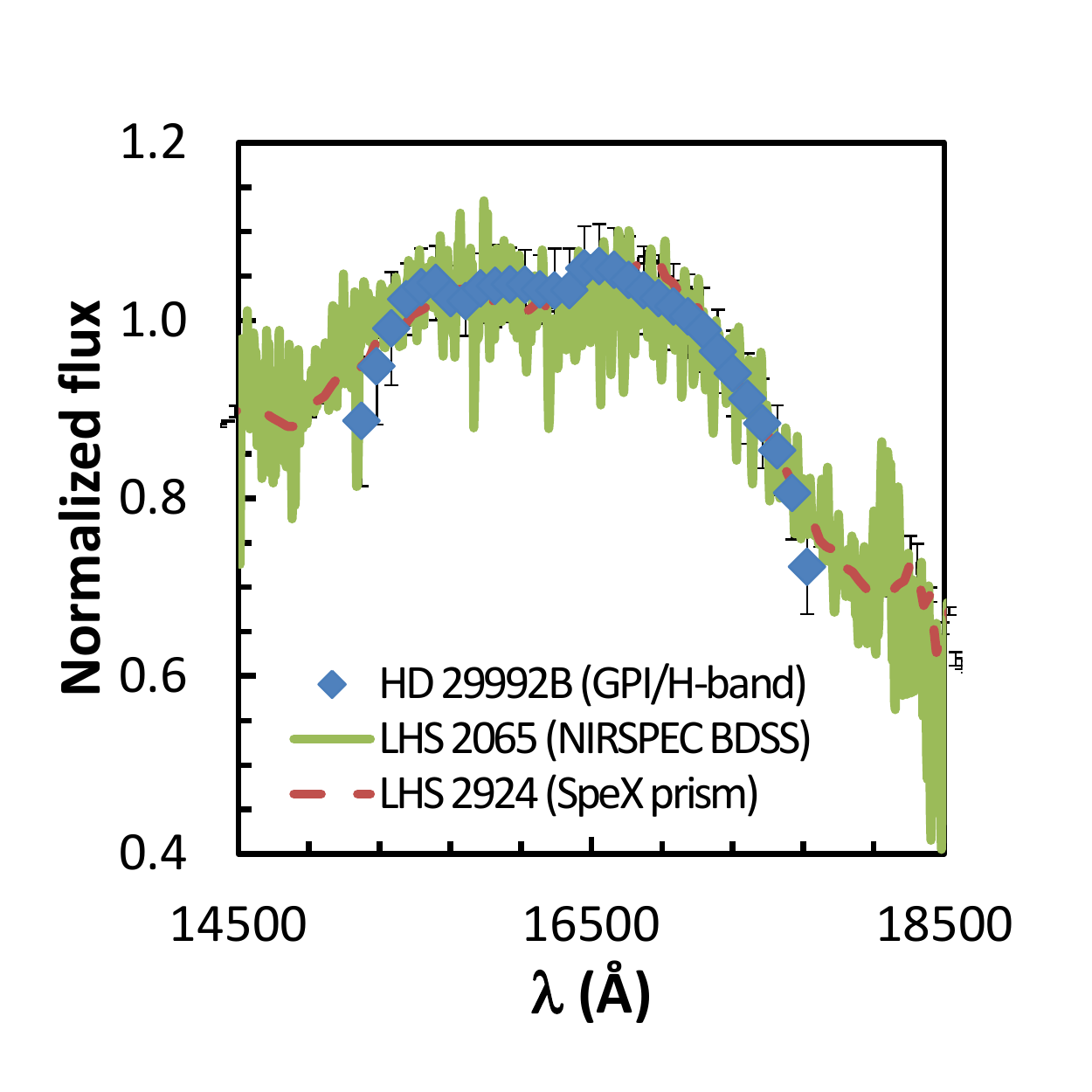}
  \includegraphics[width=0.49\linewidth,trim={0.5cm 0.8cm 0.8cm 1cm},clip]{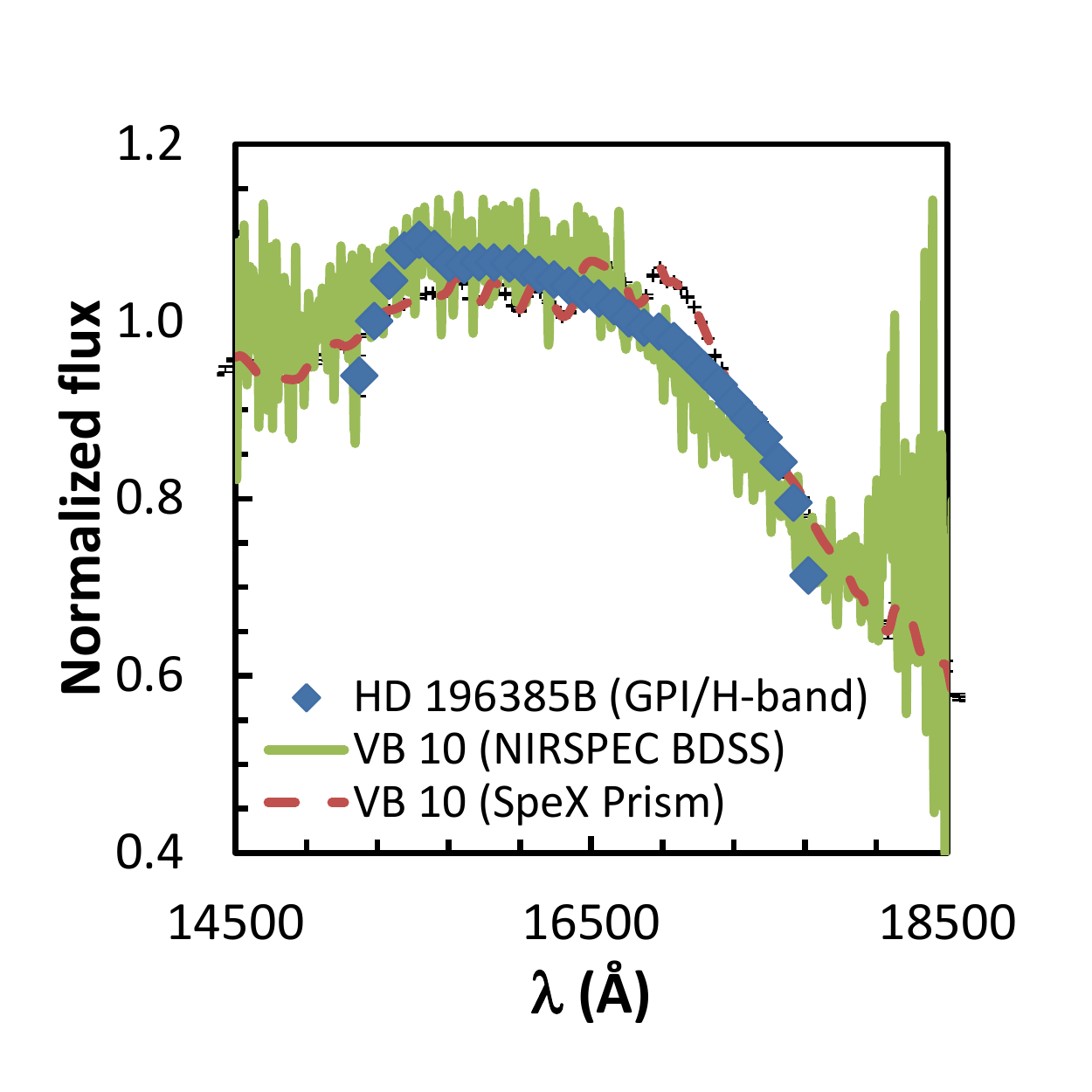}
  \caption{H-band spectra for HD 29992B (left panel), and HD 196385B (right panel) from the GPI observations (blue squares). The brown dashed line represents the best match of the SpeX Prism Spectral Libraries to the GPI data. The continuum green line corresponds to NIRSPEC BDSS library.}
  \label{fig:5}
\end{figure}

In Table~\ref{t:3}, we list the absolute magnitudes ($M_H$) assuming that the detected sources are gravitationally bound to their respective central stars, which are outside the range covered by the models of \citet{baraffe2003}. Therefore, we used the isochrones from \citet{girardi2000} for low- and intermediate-mass stars. Adopting for the companions the same ages as those for the central stars, we applied the \citet{girardi2000} models to derive the masses. Table~\ref{t:3} lists these masses, as well as the effective temperature of the detected companions to HD 29992 and HD 196385. These results are compatible with M-type stars.\\

We also extracted the H-band observed spectra of the detected sources in order to compare the overall shape of these with that of known objects in the SpeX Prism Spectral \citep{burgasser2014}, and the NIRSPEC Brown Dwarf Spectroscopic Survey \citep[NIRSPEC BDSS;][]{mclean2003} libraries. This allowed us to use another method of characterizing the detected sources, independent of the analysis with models of stellar atmospheres.\\

To perform the comparison, we wrote a routine that calculates the value of a parameter $\chi^{2}$ defined as $\chi^{2}=(F_{\lambda, {\rm GPI}}-F_{\lambda, {\rm lib}})^{2}/(\sigma_{\lambda, {\rm GPI}}^{2}+\sigma_{\lambda, {\rm lib}}^{2})$. Here, $F_{\lambda, {\rm GPI}}$ and $F_{\lambda, {\rm lib}}$ are the fluxes of the GPI and the library spectra, respectively, at the wavelength $\lambda$, with $\sigma_{\lambda, {\rm GPI}}$, and $\sigma_{\lambda, {\rm lib}}$ the errors on the fluxes. Before calculating $\chi^{2}$, the libraries spectra were degraded to the resolution of GPI convolving with a Gaussian kernel, after which they were normalized along with the GPI spectra. We cutoff the wavelengths at both ends of the GPI spectra since these regions can be affected by filter transmission and the process of microspectra extraction. Each GPI spectrum was compared to each library spectrum to derive the best match with the smallest $\chi^{2}$.\\

Figure~\ref{fig:5} shows the extracted spectra from the GPI observations, together with the best matching spectrum from the libraries cited above. For HD 29992B, the two M9V stars LHS 2924 (SpeX prim) and LHS 2965 (NIRSPEC BDSS) provided the matches with the minima residuals. In the case of HD 196385B, both libraries gave the same best match, namely the M8-type star VB 10. Therefore, the GPI H-band spectra suggest an M spectral type for both detected companions. Although some caution is advised on the results of this analysis (given the small wavelength range and spectral resolution of the data from GPI), they are compatible with those obtained using models of stellar atmospheres.\\

\subsection{Orbit fitting}
\label{orbits}

\begin{table}
\centering
  \caption{RV measurements used in the orbit fitting.}
  \label{t:7}
  \setlength{\tabcolsep}{3pt}
 \begin{tabular}{ccc|ccc}
    \hline
    \multicolumn{3}{c}{HD 29992} & \multicolumn{3}{c}{HD 196385}  \\
    BJD	& RV (ms$^{-1}$) & $\sigma$ (ms$^{-1}$) & BJD & RV (ms$^{-1}$) & $\sigma$ (ms$^{-1}$) \\ 
    \hline
2453806.51	&	$-$727	&	6	&	2453875.82	&	$-$56	&	2	\\
2453806.51	&	$-$606	&	5	&	2453875.83	&	$-$56	&	3	\\
2453806.53	&	$-$675	&	5	&	2453875.89	&	$-$60	&	3	\\
2453806.54	&	$-$512	&	5	&	2453875.90	&	$-$72	&	4	\\
2453807.53	&	$-$716	&	6	&	2453880.82	&	$-$64	&	3	\\
2453807.53	&	$-$696	&	6	&	2453880.82	&	$-$67	&	3	\\
2453989.81	&	$-$252	&	5	&	2453881.83	&	$-$31	&	3	\\
2453989.82	&	$-$317	&	5	&	2453881.83	&	$-$24	&	3	\\
2454439.73	&	272	&	6	&	2453986.75	&	$-$59	&	3	\\
2454439.73	&	186	&	8	&	2453986.76	&	$-$55	&	2	\\
2454444.71	&	124	&	5	&	2454439.54	&	$-$42	&	3	\\
2454444.71	&	78	&	5	&	2454439.55	&	$-$39	&	3	\\
2454563.47	&	629	&	5	&	2454440.55	&	$-$5	&	2	\\
2454563.48	&	452	&	5	&	2454440.56	&	$-$3	&	3	\\
2454742.84	&	297	&	12	&	2454444.53	&	$-$41	&	3	\\
2454742.85	&	175	&	8	&	2454444.54	&	$-$37	&	3	\\
2454746.89	&	190	&	5	&	2455147.56	&	11	&	2	\\
2454746.90	&	80	&	5	&	2455147.57	&	6	&	2	\\
2454759.76	&	92	&	5	&	2455148.56	&	$-$19	&	2	\\
2454759.76	&	121	&	5	&	2455148.57	&	29	&	2	\\
2454763.74	&	141	&	5	&	2455150.55	&	$-$24	&	2	\\
2454763.75	&	117	&	5	&	2455150.56	&	$-$16	&	2	\\
2454774.77	&	319	&	5	&	2455385.85	&	10	&	2	\\
2454774.77	&	428	&	5	&	2455385.86	&	4	&	2	\\
2455147.71	&	151	&	7	&	2455386.81	&	$-$10	&	2	\\
2455147.71	&	251	&	6	&	2455386.82	&	98	&	1	\\
2455172.70	&	$-$74	&	6	&	2455386.82	&	$-$14	&	2	\\
2455172.70	&	180	&	7	&	2455386.83	&	$-$5	&	2	\\
2455384.94	&	330	&	8	&	2455386.84	&	$-$18	&	2	\\
2455565.59	&	285	&	6	&	2455386.84	&	$-$7	&	2	\\
2455565.59	&	143	&	6	&	2455386.85	&	$-$4	&	2	\\
2455568.58	&	242	&	4	&	2455386.86	&	21	&	2	\\
2455568.58	&	207	&	5	&	2455724.85	&	40	&	2	\\
2455596.57	&	$-$258	&	4	&	2455724.86	&	39	&	2	\\
2455596.58	&	$-$180	&	8	&	2455726.75	&	13	&	2	\\
	&		&		&	2455726.75	&	8	&	2	\\
	&		&		&	2455726.76	&	$-$5	&	2	\\
	&		&		&	2455726.77	&	$-$4	&	2	\\
	&		&		&	2455726.78	&	$-$3	&	2	\\
	&		&		&	2455726.78	&	$-$7	&	2	\\
	&		&		&	2455726.79	&	$-$2	&	2	\\
	&		&		&	2455726.80	&	$-$10	&	2	\\
	&		&		&	2455726.81	&	$-$11	&	2	\\
	&		&		&	2455726.81	&	$-$14	&	2	\\
	&		&		&	2455726.82	&	$-$15	&	2	\\
	&		&		&	2455763.76	&	22	&	2	\\
	&		&		&	2455763.77	&	14	&	2	\\
	&		&		&	2455763.78	&	11	&	2	\\
	&		&		&	2455763.80	&	$-$1	&	2	\\
	&		&		&	2455763.81	&	2	&	2	\\
	&		&		&	2455763.82	&	$-$4	&	2	\\
	&		&		&	2455763.83	&	$-$4	&	2	\\
	&		&		&	2455763.84	&	$-$11	&	2	\\
	&		&		&	2455764.85	&	22	&	2	\\
	&		&		&	2455764.86	&	18	&	2	\\
	&		&		&	2455765.85	&	1	&	2	\\
	&		&		&	2455765.86	&	66	&	3	\\
    \hline
  \end{tabular}
\end{table}

To obtain a preliminary characterization of the orbits of the companions, we performed a join fit of the astrometry data from GPI (in Table~\ref{t:1}) and the public RV measurements available from the instrument HARPS \citep{trifonov2020}, with the RV values being  listed in Table~\ref{t:7}. We then used the python package \texttt{orbitize!}\footnote{Available at: https://github.com/sblunt/orbitize} \citep{blunt2020}, designed to compute posterior distributions of orbital elements for data covering short fractions of the orbit. \texttt{Orbitize!} implements the Bayesian rejection sampling algorithm Orbits For The Impatient \citep[OFTI;][]{blunt2017} and the \texttt{ptemcee} parallel-tempered Markov Chain Monte Carlo (MCMC) approach to sampling posterior distributions from \citet{foreman2013}. In the present study, the parallel-tempered algorithm \texttt{ptemcee} \citep{vousden2016} was applied. In addition to the RV and the astrometry data, we adopted the distances and the masses listed in Table~\ref{t:1a}.\\

We fitted six Keplerian elements (semimajor axis $a$, eccentricity $e$, inclination of the orbit $i$, argument of periastron $\omega$, position angle of nodes $\Omega$, and epoch of periastron $\tau$), plus the mass of the companion ($m_{\rm sec}$). Hereafter, we will refer to this as the ``dynamical'' mass. Using the stellar mass, the dynamical mass, and the semimajor axis values of each walker at each accepted iteration, we calculated the corresponding orbital period ($P$). \\

Tables~\ref{t:5} and~\ref{t:6} show a summary of the orbital parameters, their prior distributions, ranges, and also the results of the orbital fitting. In the following discussion we adopt the median values of the marginalized posterior distributions as being a representative estimator of each parameter's value, and the 68\% confidence intervals as the errors. For comparison, column ``Best'' lists the results for the orbit with the highest value of Log(likelihood)\footnote{The parameter Log(likelihood) is a measure of the goodness of our fits: the larger the Log(likelihood), the better the fit on the observational data.} in our runs. The corresponding Log(likelihood) for both sets of orbital parameters are also presented in Tables~\ref{t:5} and~\ref{t:6}. \\

\subsubsection{HD 29992}
\label{orbits_hd29992}

\begin{table*}
\centering
  \caption{Results for the orbits of HD 29992B.}
  \label{t:5}
 \begin{tabular}{lcccccc}
    \hline
\multicolumn{1}{c}{Orbital Element} 	&	 \multicolumn{1}{c}{Prior Range} 	&	 \multicolumn{1}{c}{Prior Distribution} 	&	 \multicolumn{1}{c}{Best} 	&	 \multicolumn{1}{c}{Median} 	&	 \multicolumn{1}{c}{68\% Confidence Range} 	&	 \multicolumn{1}{c}{95\% Confidence Range} \\
    \hline												
$a$ (au) 	&	$10^{-3}$--$10^{4}$ 	&	 Uniform in  log $a$ 	&	 4.6 	&	 4.7 	&	 4.6--4.9 	&	 4.5--5.1\\
$e$  	&	 0.0--1.0 	&	 Uniform in $e$ 	&	 0.6 	&	 0.5 	&	 0.4--0.6 	&	 0.4--0.7 \\ 
$i$ (\degr)	&	0--180	&	Uniform in sin $i$	&	153	&	117	&	20--161	&	8--173 \\
$0\leq i<90$ (\degr) 	&	 $\cdots$ 	&	 $\cdots$ 	&	$\cdots$	&	 27 	&	 13--41 	&	 6--55 \\
$90\leq i<180$ (\degr) 	&	 $\cdots$ 	&	 $\cdots$ 	&	$\cdots$	&	 153 	&	 134--168 	&	 117--174 \\
$\omega$ (\degr) 	&	 0--360 	&	 Uniform in $\omega$ 	&	 38 	&	 37 	&	 32--43 	&	 26--53 \\ 
$\Omega$ (\degr)	&	 0--360 	&	 Uniform in $\Omega$	&	151	&	61	&	 55--280	&	 30–299\\
$0\leq\Omega<180$ (\degr)  	&	 $\cdots$ 	&	$\cdots$ 	&	 $\cdots$ 	&	64	&	 44--75 	&	 26--166\\
$180\leq\Omega<360$ (\degr)  	&	 $\cdots$ 	&	$\cdots$ 	&	 $\cdots$ 	&	274	&	 263--288 	&	 256--305\\
$\tau$ 	&	 0.0--1.0 	&	 Uniform in $\tau$ 	&	 0.6 	&	 0.6 	&	 0.2--0.8 	&	0.0--1.0 \\
$m_{\rm sec}$ (M$_{\odot}$)	&	 $10^{-2}$--2  	&	 Uniform in log $m_{\rm sec}$ 	&	 0.08 	&	 0.08 	&	 0.05--0.18 	&	 0.04--0.4 \\
$P$ (yr) 	&	 $\cdots$ 	&	 $\cdots$ 	&	 7.5 	&	 7.7 	&	 7.5--8.1 	&	 7.4--8.7 \\
Log(likelihood) 	&	 $\cdots$ 	&	 $\cdots$ 	&	 -$81$ 	&	 -$6.2\times10^{4}$ 	&	 $\cdots$ 	&	 $\cdots$ \\
    \hline
  \end{tabular}
\end{table*}

\begin{figure}
  \includegraphics[width=0.95\linewidth,trim={1cm 0cm 3cm 0cm},clip]{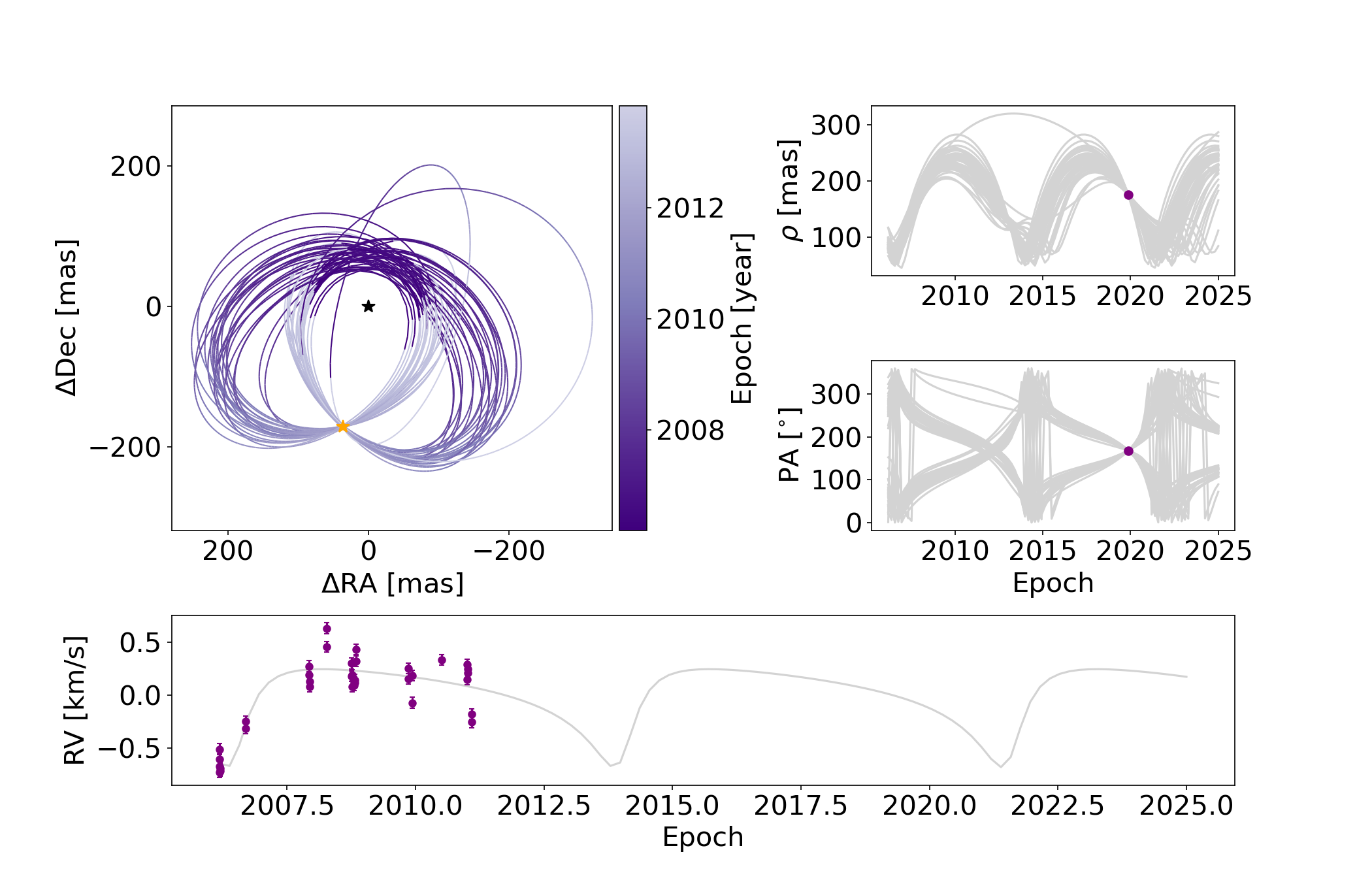}
  \caption{\emph{Left upper panel:} Plot of 50 accepted orbits randomly selected for HD 29992. The black star indicates the position of the primary component, while the orange star is the location of the companion. \emph{Right panels:} The same orbits as above (in gray) in a diagram of separation and PA vs Epoch. The purple stars are the measurements with GPI. \emph{Bottom panel:} RV time series for the primary star. The light blue dots are the RV measurements from HARPS. The solid gray line is the best fit found with \texttt{orbitize!}.}
  \label{fig:7}
  
\end{figure}
\begin{figure}
  \includegraphics[width=0.95\linewidth,trim={0cm 0cm 0cm 0cm},clip]{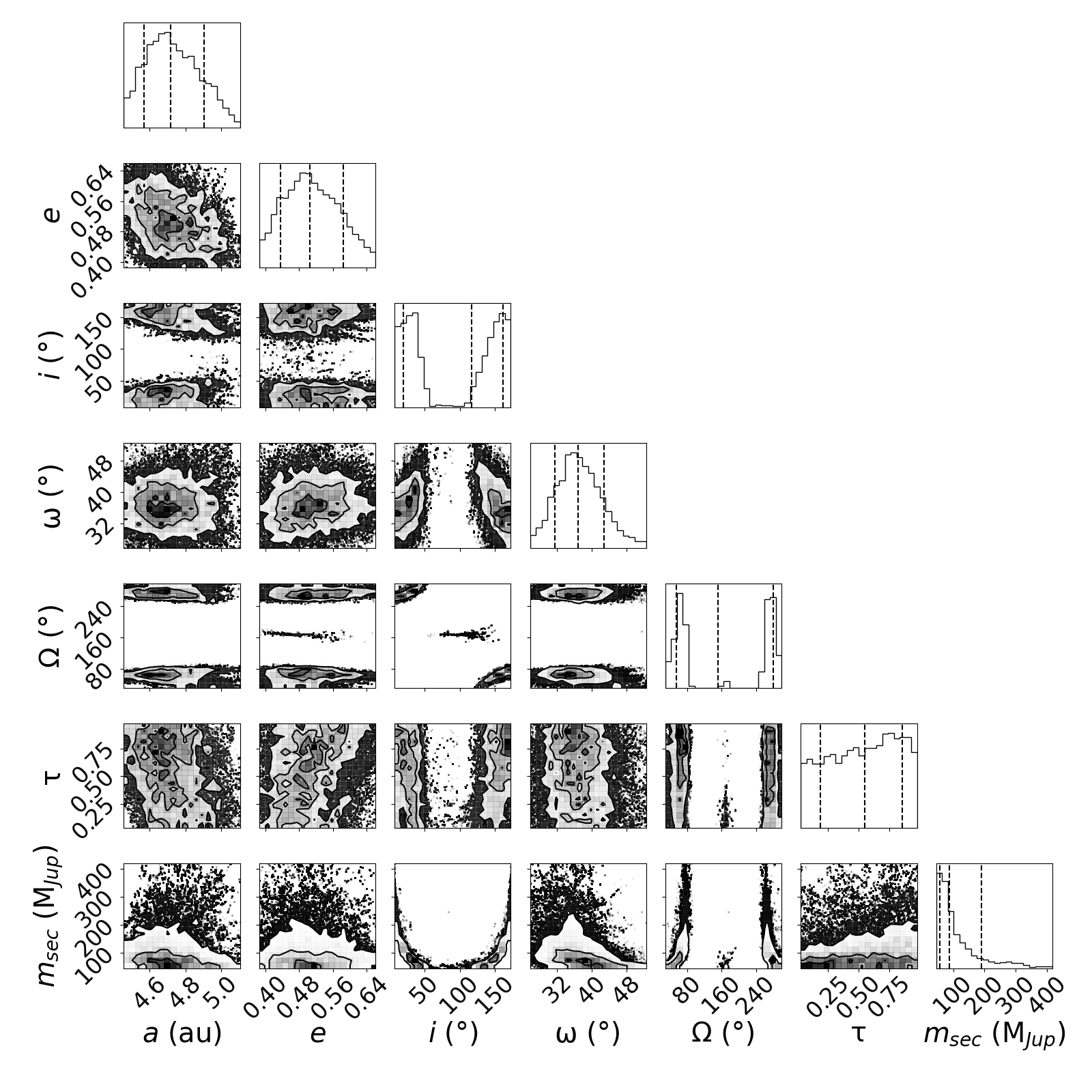}
  \caption{Posterior and joint distributions built using $10^6$ accepted orbits for HD 29992B. The vertical dashed lines indicate the 16th, 50th, and 84th percentile values. For clarity, the panels show only 95\% of the samples.}
  \label{fig:6}
\end{figure}

For HD 29992 we performed several initial tests varying the number of required accepted orbits and burn-in steps. Finally, we required 1000000 accepted orbits for building the posterior distributions. The MCMC sampler was run for 1000 steps using 1000 walkers at each of the 20 temperatures after 26500 steps of burn-in, in order to allow the walkers to converge. Median values together with the 68\% and 95\% confidence ranges for each parameter are listed in Table~\ref{t:5}, and Figure~\ref{fig:7} shows a plot of 50 accepted orbits randomly selected.\\

The median for the distribution of the semimajor axis is $4.7^{+0.2}_{-0.1}$ au with the corresponding median for the period distribution being $7.7^{+0.4}_{-0.2}$ yr. Our fit tightly constrains the semimajor axis for the possible orbits. This value is close to the projected separation measured in the images ($\sim5$ au). For the eccentricity, the fit seemed to favor orbits with relatively high eccentricities, with the median being $0.5^{+0.1}_{-0.1}$ and values as high as $e\sim0.7$ at the 95\% confidence level. \\

The orbit inclination ($i$) is not very well constrained by our fit. The distribution in Figure~\ref{fig:6} reveals two separate clusters of accepted orbits around $i\sim27\degr$ and $i\sim153\degr$. In this case, the median is not a good reference value. Table~\ref{t:5} lists the medians and confidence ranges for orbits with $0\leq i<90\degr$ and $90\leq i<180\degr$. Both sets of orbits are closer to the plane of the sky than to an edge-on configuration ($i\sim90\degr$) and produce RV curves of similar characteristics. The difference lies in that for orbits of $0\leq i<90\degr$ the companion orbits counterclockwise in the sky (North-up, East-left), while for inclinations of $90<i\leq 180\degr$ the motion is clockwise. In either case, neither of the orbits is coplanar with the equatorial plane of the central star ($i_{\star}\sim65.3\degr$). However, low inclination orbits are consistent with a low-mass star companion. If this was not the case, then larger amplitudes for the RVs would have been expected.\\

The argument of periastron ($\omega$) is relatively constrained with a median value of $37\degr^{+6}_{-5}$. On the other hand, the position angle of the nodes ($\Omega$) is poorly constrained by our fit and reveals a similar behavior as the inclination angle, with a large number of orbits found with $\Omega\sim64\degr$ and $\Omega\sim274\degr$ (which is apparent from the random orbits plotted in Figure~\ref{fig:7}). Table~\ref{t:5} lists the medians and confidence ranges for orbits with $0\leq\Omega<180\degr$ and $180\leq\Omega<360\degr$. This may have been due to the fact that only one separation for a given date is available ($\Omega$ has no effect on the resulting RV curve), making hard to know the orientation of the line of nodes on the plane of the sky. Additional measurements of separation and PA at different epochs would help to constrain better the range of possible $\Omega$s.\\

For the dynamical mass of the companion, the median value is $0.08^{+0.10}_{-0.03}$ M$_{\odot}$. The fit favored smaller masses than the one estimated using stellar atmosphere models ($\sim0.19$ M$_{\odot}$) and was closer to the brown dwarf regime. However, the distribution of $m_{\rm sec}$ in Figure~\ref{fig:6} contains many accepted orbits for objects with dynamical mass $>0.1$ M$_{\odot}$ and values as high as $\sim0.4$ M$_{\odot}$ at the 95\% confidence level.\\

In general, we found a good agreement between the parameters of the orbit constructed from the median values, and the orbit with the highest probability in our run (i.e. highest Log(likelihood)). The exceptions are the inclination $i$ and the position angle of nodes $\Omega$ that show a bi-modal behavior in their distributions. The median value is not a good reference value in these cases and the orbit constructed from the median values has a significantly lower Log(likelihood) than our orbit with the highest probability. Moreover, the orbital fits are consistent with the astrometry datapoints of separation, position angle (right panels in Figure~\ref{fig:7}) and radial velocity (lower panel in Figure~\ref{fig:7}). The relatively small amplitude observed in the RV data may have favored the orbits of objects with smaller dynamic masses than those estimated using atmospheric models. On the other hand, the two clusters of orbits found around $i\sim27\degr$ and $i\sim153\degr$ agree better with the presence of a low-mass star companion. Figure~\ref{fig:7} shows that the RV grows from $\sim-0.7$ kms$^{-1}$ to $\sim+0.4-0.6$ kms$^{-1}$ at around 2007.5 before slowly starting to decrease again at later dates, but never reaching the minimum values initially detected. Therefore, it is not possible from these data to infer if the RV could have smaller values (these data extend to about 60\% of the median orbital period). New observations of RV would be required to be able to put stronger constrains on the dynamical mass.\\
 
\subsubsection{HD 196385}
\label{orbit_hd196385}

\begin{table*}
\centering
  \caption{Results for the orbits of HD 196385B.}
  \label{t:6}
 \begin{tabular}{lcccccc}
    \hline
    \multicolumn{1}{c}{Orbital Element} & \multicolumn{1}{c}{Prior Range} & \multicolumn{1}{c}{Prior Distribution} & \multicolumn{1}{c}{Best} &\multicolumn{1}{c}{Median} & \multicolumn{1}{c}{68\% Confidence Range} & \multicolumn{1}{c}{95\% Confidence Range}\\
    \hline
$a$ (au) 	&	$10^{-3}$--$10^{4}$ 	&	 Uniform in  log $a$ 	&	33	&	 36 	&	 33--40 	&	 30--45 \\
$e$  	&	 0.0--1.0 	&	 Uniform in $e$ 	&	0.6	&	 0.5 	&	 0.4--0.6 	&	 0.3--0.8 \\ 
$i$ (\degr) 	&	 0--180 	&	 Uniform in sin $i$ 	&	46	&	 66 	&	 52--83 	&	 36--105 \\
$\omega$ (\degr) 	&	 0--360 	&	 Uniform in $\omega$ 	&	259	&	231 	&	 211--251 	&	 193--278 \\ 
$\Omega$ (\degr) 	&	 0--360 	&	 Uniform in $\Omega$ 	&	147	&	193 	&	 171--212 	&	 144--232 \\
$\tau$ 	&	 0.0--1.0 	&	 Uniform in $\tau$ 	&	0.6	&	0.5 	&	0.4--0.5 	&	0.3--0.7 \\
$m_{\rm sec}$  (M$_{\odot}$)	&	 $10^{-6}$--2  	&	 Uniform in log $m_{\rm sec}$ 	&	0.26	&	 0.30 	&	 0.25--0.36 	&	 0.20--0.44 \\
$P$ (yr) 	&	 $\cdots$ 	&	 $\cdots$ 	&	151	&	 161 	&	 141--186 	&	 124--223 \\
Log(likelihood)	&	 $\cdots$ 	&	 $\cdots$ 	&	137	&	53	&	 $\cdots$ 	&	 $\cdots$ \\
    \hline
  \end{tabular}
\end{table*}

\begin{figure}
  \includegraphics[width=0.95\linewidth,trim={0cm 0cm 0cm 0cm},clip]{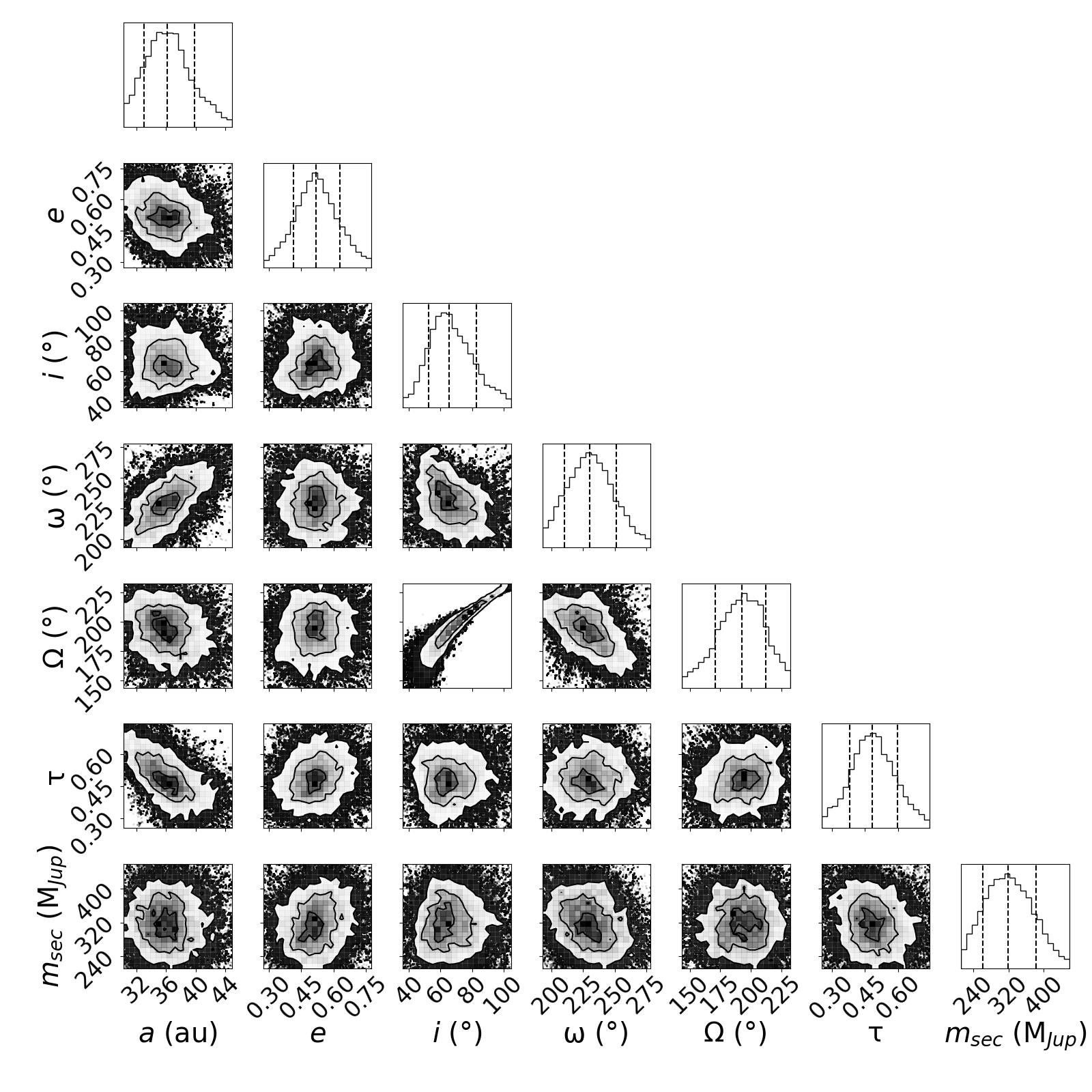}
  \caption{Posterior and joint distributions for the fitted parameters for HD 196385B. The vertical dashed lines indicate the 16th, 50th, and 84th percentile values. The one-dimensional distributions of each element is shown on the diagonal. For clarity, the panels show only 95\% of the samples.}
  \label{fig:8}
\end{figure}

\begin{figure}
  \includegraphics[width=1.\linewidth,trim={1cm 0cm 3cm 0cm},clip]{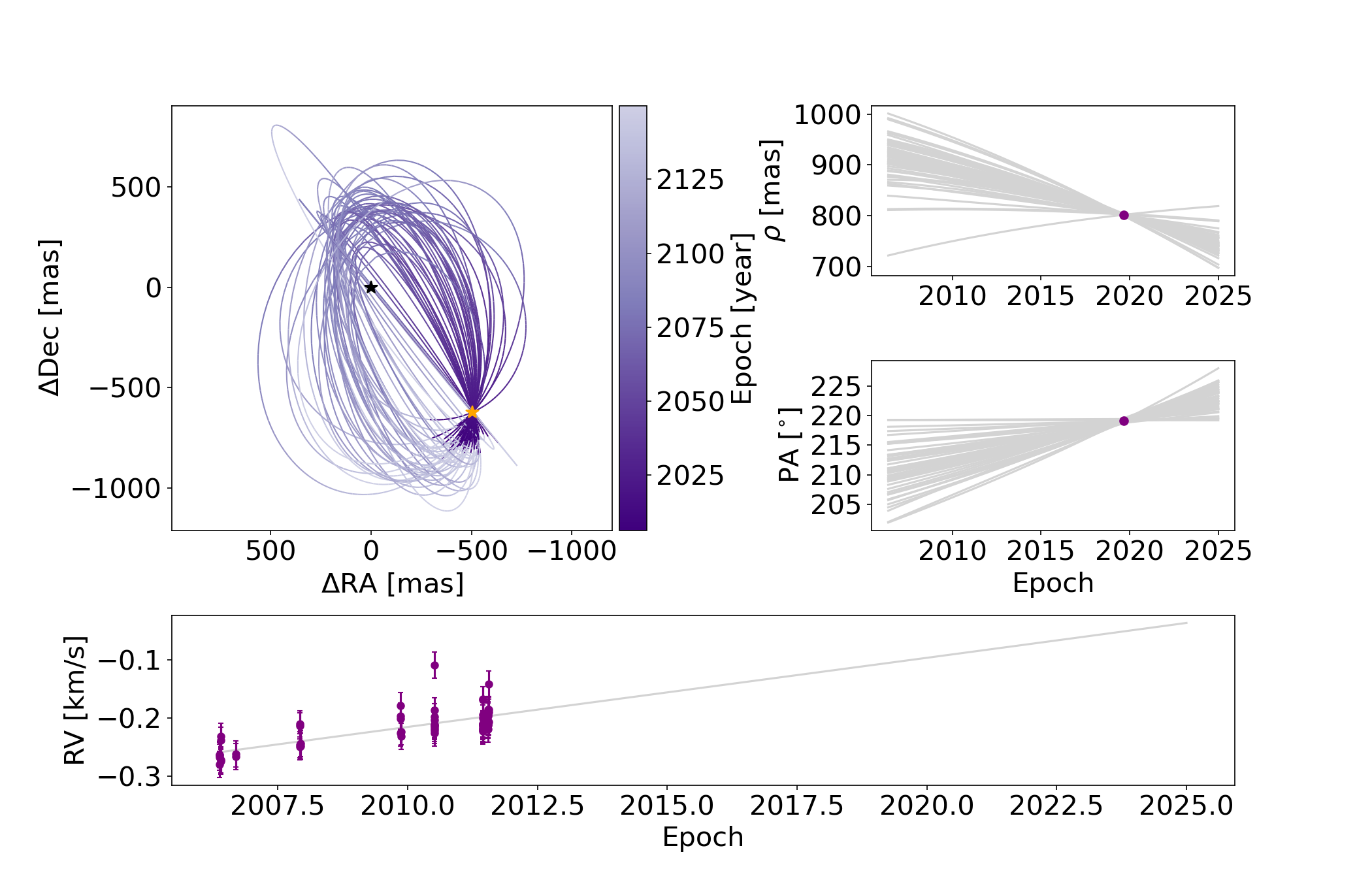}
  \caption{\emph{Left upper panel:} Plot of 50 accepted orbits randomly selected. The black and orange stars indicate the position of the primary component and the companion, respectively. \emph{Right panels:} The same orbits as above (in gray) in a diagram of separation and PA vs time. The purple stars represent the measurements with GPI. \emph{Bottom panel:} RV time series for the primary star. The light blue dots are the RV measurements from HARPS. The solid gray line is the best fit from \texttt{orbitize!}}
  \label{fig:9}
\end{figure}

For HD 196385, we ran an initial test to obtain 1000000 accepted orbits in the same way as that described above for HD 29992. Then, we used the results from this run to define the starting position of the 1000 walkers to obtain, once again, 1000000 accepted orbits, but now with 100 burn-in steps. The median values from the first run, did not differ significantly from those obtained in the second run. We proceeded in this way to improve the fit over the RV data. The results are presented in Table~\ref{t:6}. Posteriors and joint distributions are shown in Figure~\ref{fig:8}, while Figure~\ref{fig:9} shows a plot of 50 random accepted orbits.\\ 

We found a median semimajor axis of 36 au, with 68\% confidence between 33 and 40 au and a corresponding period of 161 years, with 68\% confidence between 141 and 186 years. The measured projected separation from the images ($\sim38$ au), albeit larger than the median value, it is within the 68\% confidence level. The median eccentricity is $0.5^{+0.1}_{-0.1}$ with values as high as 0.8 at the 95\% confidence level, suggesting rather highly eccentric orbits for HD 196385B.\\

For the inclination, we obtained a median value of $i\sim66\degr$. The distribution of $i$ in Figure~\ref{fig:8} revealed a relatively broad range of inclinations with values between $\sim52\degr$ and $\sim83\degr$ at 68\% confidence level. It is worth mentioning, however, that this range does not include edge-on orbits, and suggests that the orbit of HD 196385B would not be coplanar with the equatorial plane of the star ($i_{\star}\sim17\degr$). The argument of periastron $\omega$ and the position angle of the nodes $\Omega$ are both poorly constrained by our fit. For $\omega$ and $\Omega$, the medians found are $231\degr^{+20}_{-20}$ and $193\degr^{+19}_{-22}$, respectively. Finally, the median dynamical mass found of $0.30^{+0.06}_{-0.05}$ M$_{\odot}$ is in good agreement with the mass derived from the photometry and the stellar atmosphere models.\\

For HD 196385 Figure~\ref{fig:9} shows that the orbital fits are consistent with the measurements of separation, position angle (right panels in Figure~\ref{fig:9}) and measured radial velocity (lower panel in Figure~\ref{fig:9}). The median semimajor axis is in good agreement with the projected separation measured in the images as well as the median dynamical mass in comparison with the value derived from the atmosphere models. The parameters for the best orbit, and those corresponding to the median orbit, are within the 68\% confidence range. However, the position angle of nodes $\Omega$, the argument of periastron $\omega$, and the inclination $i$ have a larger discrepancy. For $\Omega$ this is expected because this parameter is not well constrained with only one measurement of separation and position angle. Radial velocity measurements show a gradual increase in velocity. However, the observations do not extend far enough in time\footnote{RV data cover $\sim$3\% of the 161-year period, or 5.2 years} to determine a maximum or minimum value that would help place stronger constraints on $\omega$ and $i$. Given the long period inferred for this system, further astrometric measurements of separation and position angle would be particularly useful in narrowing down the set of possible orbits for HD 196385B.\\

\section{Summary and conclusions}
\label{sec:conc}

We report the detection of two point-like sources at angular separations of $\rho\sim0.18\arcsec$ and $\rho\sim0.80\arcsec$ from the stars HD 29992 and HD 196385, respectively. The combined analysis of the new GPI observations and previous high spatial resolution images from \citet{ehrenreich2010} suggests that the source detected close to HD 29992 is gravitationally bound to the star. The measured projected separation ($\sim5$ au) and estimated mass ($\sim0.2$ M$_{\odot}$) determined using models of stellar atmosphere are in good agreement with the characteristics predicted by \citet{borgniet2016} from their analysis of RV measurements of the star. Moreover, we can discard the presence of additional companions with $m>75$ M$_{\rm Jup}$ between $\rho\sim0.3\arcsec$ and $\rho\sim1.5\arcsec$ from HD 29992. \\

For HD 196385, we used statistical arguments to evaluate the companionship of the detected source, and found that about $\sim0.5$ objects would be expected in the GPI images. This suggests that the chances of a fortuitous alignment with a background star are rather low and that the detected source may be gravitationally bound to HD 196385. In that case, the companion would have a mass of $\sim0.3$ M$_{\odot}$ according to the models of stellar atmospheres. In addition, the contrast achieved in the individual datacubes allows us to rule out any potential companions with $m>75$ M$_{\rm Jup}$ between $\rho\sim0.2\arcsec$ and $\rho\sim1.3\arcsec$. \\

We extracted and compared the observed H-band spectra of the detected sources with those of previously known objects. In both cases, we found that the best matches for the observed spectra were those of late M-type stars. This is roughly in agreement with the inferred masses using theoretical models of stellar atmospheres.\\

Using the python package \texttt{orbitize!} we performed a joint fit of the new GPI astrometry data and previously existing RV measurements to infer the most likely orbits for HD 29992B and HD 196385B. For both systems, we found a good agreement between the projected separation measured in the images and the median semimajor axis of the posterior distributions. For HD 196385B, the median dynamical mass is also in good agreement with the estimated mass using the stellar atmosphere models. For HD 29992B, the orbital solutions favored a dynamic mass (median of 0.08 M$_{\odot}$) closer to the brown dwarf regime than that estimated using models of stellar atmospheres (0.2 M$_{\odot}$). The small dynamical mass is hard to reconcile with the high brightness observed in the GPI images (assuming that the source is at the distance of HD 29992). One possible explanation for this is that the inclination ($i$) of the orbit is low. As the posterior distribution of $i$ for HD 29992B shows two clusters of accepted orbits of $i\sim27\degr$ and $i\sim153\degr$, this can explain the relatively low amplitude observed in the RV curve and it is compatible with the mass estimated using stellar atmosphere models and the prediction from \citet{borgniet2016}.\\

Our analysis suggests that HD 29992 and HD 196385 could be two binary systems with an M-type secondary component orbiting at separations of $\sim5$ and $\sim38$ au, respectively, in relatively low inclination orbits. However, new images are required to measure the proper motion of the detected sources and to definitely confirm whether they are co-moving with the central star. In addition, these images would help to put stronger constraints on the characteristics of both systems, in particular in the case of HD 29992.\\

\section*{Acknowledgements}

Based on observations obtained at the international Gemini Observatory, a program of NOIRLab, which is managed by the Association of Universities for Research in Astronomy (AURA) under a cooperative agreement with the National Science Foundation. on behalf of the Gemini Observatory partnership: the National Science Foundation (United States), National Research Council (Canada), Agencia Nacional de Investigaci\'{o}n y Desarrollo (Chile), Ministerio de Ciencia, Tecnolog\'{i}a e Innovaci\'{o}n (Argentina), Minist\'{e}rio da Ci\^{e}ncia, Tecnologia, Inova\c{c}\~{o}es e Comunica\c{c}\~{o}es (Brazil), and Korea Astronomy and Space Science Institute (Republic of Korea). \\
This work has made use of data from the European Space Agency (ESA) mission Gaia (https://www.cosmos.esa.int/gaia), processed by the Gaia Data Processing and Analysis Consortium (DPAC, https://www.cosmos.esa.int/web/gaia/dpac/consortium). Funding for the DPAC has been provided by national institutions, in particular the institutions participating in the Gaia Multilateral Agreement. This paper includes data collected by the TESS mission that are publicly available from the Mikulski Archive for Space Telescopes (MAST). We acknowledge the use of public TESS data from pipelines at the TESS Science Office and at the TESS Science Processing Operations Center. Resources supporting this work were provided by the NASA High-End Computing (HEC) Program through the NASA Advanced Supercomputing (NAS) Division at Ames Research Center for the production of the SPOC data products. Funding for the TESS mission is provided by NASA’s Science Mission Directorate.\\
The authors wish to thank the referee for the valuable suggestions that helped to improve the paper, and also to Sarah Blunt for her assistance in using the package \texttt{orbitize!}.\\

\section*{Data Availability}
 
The data in this article are available from the Gemini Observatory Archive (https://archive.gemini.edu) with program ID GS-2019B-Q-107. The data products generated from the raw data are available upon request to the author.



\bibliographystyle{mnras}
\bibliography{biblhgarcia} 




\appendix

\section{Calculation of the inclinations from TESS data}
\label{sec:ap-A}

We estimated the inclination ($i_{\star}$) of HD 29992 and HD 196385 using the classical equation:

\begin{equation}{\label{eq1}}
    i_{\star} = sin^{-1}\frac{P_{rot} \times vsin(i_{\star})}{2\pi R_{\star}}
\end{equation}

\noindent where $P_{rot}$ is the rotation period, $vsin(i_{\star})$ the projected rotational velocity, and $R_{\star}$ the radius of the star. 

\begin{figure*}
\centering
\includegraphics[width=.48\textwidth]{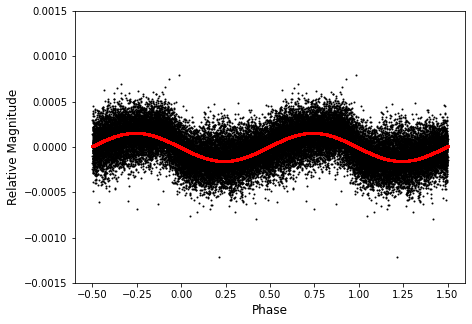}
\includegraphics[width=.51\textwidth]{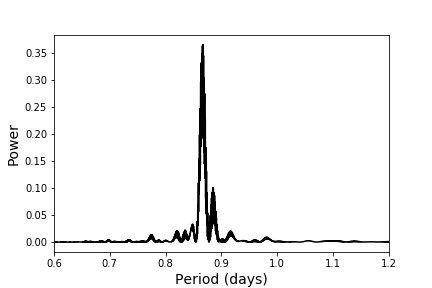}
\caption{Left: TESS phase-folded light curve of HD 29992. Black crosses indicate the un-binned TESS PDC\_SAP data, and the red continuous line points out the model corresponding to the best period detected. Right: LS periodogram.
 \label{fig.TESS.HD29992}}
\end{figure*}

\begin{figure*}
\centering
\includegraphics[width=.48\textwidth]{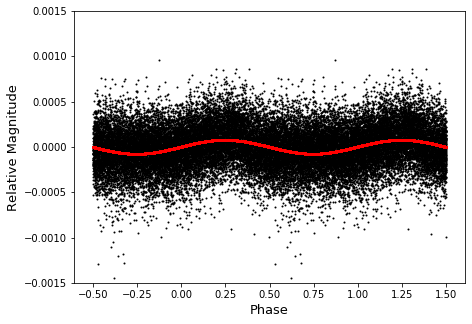}
\includegraphics[width=.51\textwidth]{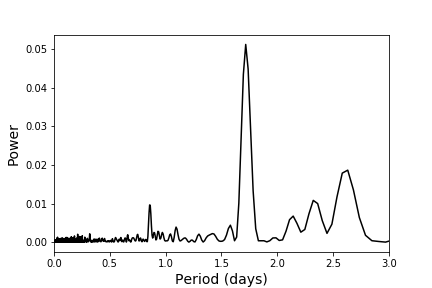}
\caption{Left: TESS phase-folded light curve of HD 196385. Black crosses indicate the un-binned TESS PDC\_SAP data, and the red continuous line points out the model corresponding to the best period detected. Right: LS periodogram.
 \label{fig.TESS.HD196385}}
\end{figure*}

Rotation periods were determined from the 2-minute cadence data provided by the Transiting Exoplanet Survey Satellite \citep[TESS;][]{Ricker2015}. We used the tools available in the \textit{Lightkurve} Python package \citep{Lightkurve2018} to retrieve and analyze the \textit{Presearch Data Conditioning Simple Aperture Photometry} (PDC\_SAP), processed with the TESS Science Processing Operations Center (SPOC) pipeline \citep{jenkins2016}. 
HD 29992 (TIC 77288515) was observed by TESS in sectors 4-5  (18 October to 11 December 2018) and 31–32 (21 October to 17 December 2020). However, given that all data from sector 4 and those at the end of sector 5 between 2458464.0 and 2458465.0 BJD (Barycentric Julian Date) were severely affected by systematics, we did not use them for further analysis. For HD 196385 (TIC 212296912), TESS collected data in sectors 1 (25 July to 22 August 2018) and 27 (4 July to 30 July 2020). In this case, we did not consider either the data between 2458347.2 and 2458349.4 or between 2458353.0 and 2458354.0 BJD due to the presence of notorious uncorrected systematics and we discarded all sector 27 data because of the inclusion of flux coming from another object in the light curve. We searched for periodic modulations in the time-series observations by running a Lomb-Scargle (LS) periodogram \citep{scargle1982} provided by the \textit{astropy} package \citep{astropy}. For HD 29992, we found $P_{rot}=0.8669\pm 0.0013$ days with an amplitude\footnote{The amplitude of the variation was computed as $\sqrt{A_{sin}^2+A_{cos}^2}$, where $A_{sin}$ and $A_{cos}$ are the amplitudes of the sine and cosine terms of the best model found by the LS periodogram.} of, $A=2.17\times 10^{-4}$ mag; whilst for HD 196385, the period detected was $P_{rot}=1.7178\pm 0.0361$ days with an amplitude $A=7.7\times 10^{-5}$ mag. TESS phase-folded data and the LS periodograms for both stars are shown in Figure~\ref{fig.TESS.HD29992} and Figure~\ref{fig.TESS.HD196385}. Finally, we superimposed the TESS fields and the adopted apertures for HD 29992 and HD 196385 over the location of nearby \textit{GAIA} DR2 sources. The fact that no \textit{GAIA} sources, with a difference in magnitude relative to the star of interest smaller than 8, fell inside or in the surroundings of the selected apertures would indicate that the origin of the signal effectively comes from the sources under study. Moreover, none of the \textit{Data Release Notes} of the sectors analyzed reported additional periodic systematics that might mimic the peaks found by LS. All these considerations lead us to suggest with high confidence that the periodic signals found have an astrophysical origin and come from the targets analyzed in this work. As far as we are aware, this would be the first time that photometric rotation periods have been reported for these stars.
In order to check whether the rotation rates are consistent with the ages presented in Table~\ref{t:1a}, we used gyrochronology \citep{barnes2007}. This technique relates the stellar rotation period $P_{rot}$ to the age, $t$, and the color $(B-V)_{0}$ by the formula $P_{rot}=t^n \times a[(B-V)_{0}-c]^b$ where $a$, $b$, and $c$ are known as the gyrochronology parameters. Recently, \cite{chowdhury2018} analyzed 15106 A - K stars observed by the Kepler satellite and showed that this relationship can be extended to colors $(B-V)_{0}$ down to 0.2 by considering different values of the parameter $c$. In this work, we adopted $a=0.77$ and $b=0.55$ from \cite{meibom2009}, $n=0.5$ following \cite{skumanich1972} and $c=0.36$ and 0.3 for HD 29992 and  HD 196385, respectively, as arbitrarily selected in \cite{chowdhury2018}. Colors $(B-V)_{0}$ were calculated as the difference between the values of $B$ and $V$ magnitudes extracted from \textit{SIMBAD}. We obtained $(B-V)_{0}=0.37$ for HD 29992 and $(B-V)_{0}=0.307$ for HD 196385. Using all this information, we computed gyrochronology ages of 1.17 and 0.2 Gyr for HD 196385 and HD 29992, respectively. For the first star, this value is, within errors, well in agreement with the age of Table~\ref{t:1a}. For HD 29992, although the age obtained from
gyrochronolgy is not within the uncertainty from that derived by \citet{casagrande2011}, it differs by only 1.5 Gyr.

Stellar radii of $R_{\star}=1.89\pm 1.80 ~R_{\odot}$ for HD 29992 and $R_{\star}=1.51\pm 1.46 ~R_{\odot}$ for HD 196385  were extracted from the \textit{GAIA} DR2 catalog \citep{gaia2016, gaia2018}, and rotational velocities of $vsin(i_{\star})=100$~kms$^{-1}$ and $vsin(i_{\star})=13$~kms$^{-1}$ were taken from the work of \citet{borgniet2016}. We introduced these values and the rotation periods estimated from the TESS data in Eq. \ref{eq1} and obtained an inclination of $i_{\star}=65.3\degr$ for HD 29992 and $i_{\star}=16.9\degr$ for HD 196385.

\bsp	
\label{lastpage}
\end{document}